\documentclass[%
 reprint, longbibliography,
superscriptaddress,
 amsmath,amssymb,
aps,
prb,
]{revtex4-2}

\usepackage{graphicx}
\usepackage{dcolumn}
\usepackage{bm}
\usepackage[colorlinks=true, linkcolor=blue, citecolor=blue]{hyperref}

\usepackage{pdfpages}
\usepackage{pgffor}
\makeatletter
\AtBeginDocument{\let\LS@rot\@undefined}
\makeatother

\begin{document}
\title{Tuning proximity-induced spin-orbit coupling in graphene/WSe$_{2}$ heterostructures}


\author{Tobias Rockinger}
\affiliation{Institute of Experimental and Applied Physics, University of Regensburg, 93040 Regensburg, Germany}
\author{Bálint Szentpéteri}
\affiliation{Department of Physics, Budapest University of Technology and Economics, Műegyetem rkp. 3., 1111 Budapest, Hungary}
\affiliation{MTA-BME Correlated van der Waals Structures Momentum Research Group, Műegyetem rkp. 3., 1111 Budapest, Hungary}
\author{Szabolcs Csonka}
\affiliation{Department of Physics, Budapest University of Technology and Economics, Műegyetem rkp. 3., 1111 Budapest, Hungary}
\affiliation{MTA-BME Superconducting Nanoelectronics Momentum Research Group, Műegyetem rkp. 3., 1111 Budapest, Hungary}
\author{Marina Marocko}
\affiliation{Institute of Experimental and Applied Physics, University of Regensburg, 93040 Regensburg, Germany}
\author{Julia Amann}
\affiliation{Institute of Experimental and Applied Physics, University of Regensburg, 93040 Regensburg, Germany}
\author{Ziyang Gan}
\affiliation{Institute of Physical Chemistry, Friedrich Schiller University Jena, 07743 Jena, Germany}
\author{Antony George}
\affiliation{Institute of Physical Chemistry, Friedrich Schiller University Jena, 07743 Jena, Germany}
\author{Andrey Turchanin}
\affiliation{Institute of Physical Chemistry, Friedrich Schiller University Jena, 07743 Jena, Germany}
\author{Kenji Watanabe}
\affiliation{Research Center for Electronic and Optical Materials, National Institute for Materials Science, 1-1 Namiki, Tsukuba 305-0044, Japan}
\author{Takashi Taniguchi}
\affiliation{Research Center for Materials Nanoarchitectonics, National Institute for Materials Science,  1-1 Namiki, Tsukuba 305-0044, Japan}
\author{Dieter Weiss}
\affiliation{Institute of Experimental and Applied Physics, University of Regensburg, 93040 Regensburg, Germany}
\author{Péter Makk}
\affiliation{Department of Physics, Budapest University of Technology and Economics, Műegyetem rkp. 3., 1111 Budapest, Hungary}
\affiliation{MTA-BME Correlated van der Waals Structures Momentum Research Group, Műegyetem rkp. 3., 1111 Budapest, Hungary}
\author{Jonathan Eroms}
\email{jonathan.eroms@ur.de}
\affiliation{Institute of Experimental and Applied Physics, University of Regensburg, 93040 Regensburg, Germany}

\date{\today}

\begin{abstract}
Recently, proximity-induced spin-orbit coupling (SOC) has been observed in
heterostructures consisting of monolayer graphene (ML-G) and transition metal dichalcogenides (TMDCs) such as  WSe$_{2}$. Successful tuning of  SOC in graphene/WSe$_{2}$ heterostructures by applying mechanical pressure and electric fields was also demonstrated in previous studies. In addition, theoretical calculations predicted a strong dependence of the proximity-induced SOC on the twist angle between graphene and TMDC. Here, we put these predictions to experimental test in 
ML-G/ML-WSe$_{2}$/hBN-heterostructures, where the twist angle is determined by aligning fractured edges, and by crystallographic etching of graphene. By performing weak anti-localization measurements, we determine the strength of the Rasbha-type SOC ($\lambda_\mathrm{R}$) and the valley-Zeeman-type SOC ($\lambda_\mathrm{VZ}$). Our experiments confirm a strong twist angle dependence of the proximity-induced SOC in agreement with theoretical predictions. Finally, we demonstrate the tunability of the SOC strength via mechanical pressure, which is in agreement with earlier findings.
\end{abstract}

\maketitle


\section{\label{sec:level1}Introduction\protect\\}
Graphene is ideally suited for spintronic applications due to its low intrinsic spin-orbit coupling (SOC) and therefore long spin lifetimes \cite{Gmitra2,HanW}. However, pristine graphene is not well-adapted for generating or manipulating spin currents directly due to its weak intrinsic SOC \cite{Gmitra1,Gmitra3}. This limitation can be overcome by enhancing SOC in graphene, enabling the generation and manipulation of spin currents  \cite{Kaloni2014,Gmitra3,Avsar,ZheWang,Zihlmann,Voelkl,Wakamura,Cummings,Benitez1}. 
A particularly promising method involves coupling graphene to transition metal dichalcogenides (TMDCs) such as WSe$_{2}$, which has been shown to induce strong SOC in graphene \cite{ZheWang,Zihlmann,Voelkl,Wakamura}. It has also been demonstrated that proximity-induced SOC can be tuned in situ by gate voltages \cite{Khoo.2017,Gmitra.2017,Island2019,Wang2019,Amann,Dulisch2025} or by applying high pressure \cite{Balint,Kedves2023,BalintPRB2025}. To gain more control over proximity-induced SOC and, most importantly, to achieve better reproducibility, it is necessary to consider the angular orientation between monolayer graphene (ML-G) and monolayer WSe$_{2}$ (ML-WSe$_{2}$), which has been predicted to significantly affect proximity-induced SOC \cite{YangLi,DavidA,NaimerT,ZollnerK} and shown to modify the spin texture in a precession experiment \cite{Yang2024}. Therefore, we investigated the angular dependence of proximity-induced SOC in ML-G/ML-WSe$_{2}$/hBN heterostructures by fabricating samples with well-defined rotation angles and conducting magnetotransport measurements. The weak anti-localization (WAL) effect, which occurs at small magnetic fields and is highly sensitive to SOC \cite{McCannFalko,ZheWang}, was used to determine the amount of proximity-induced SOC \cite{Zihlmann,Cummings,Ilic2019}. This allows us not only to show the strong twist angle dependence of the induced SOC, but also to show that samples with similar SOC strength can be reproducibly fabricated when using the same twist angle. We also investigated the pressure dependence of the proximity-induced SOC and confirm the results from Ref.~\cite{Balint}, where a significant pressure dependence of the Rashba-type SOC was shown.

\section{\label{sec:2}Tuning proximity-induced SOC using twist angle dependence\protect\\}
A total of six samples were investigated and  the main findings are presented here, whereas additional data are shown in the Supplementary Material \cite{SM}.
\subsection{\label{subsec:level1}Sample Fabrication\protect\\}
\begin{figure}[h] 
	\centering
	\includegraphics[width=8cm]{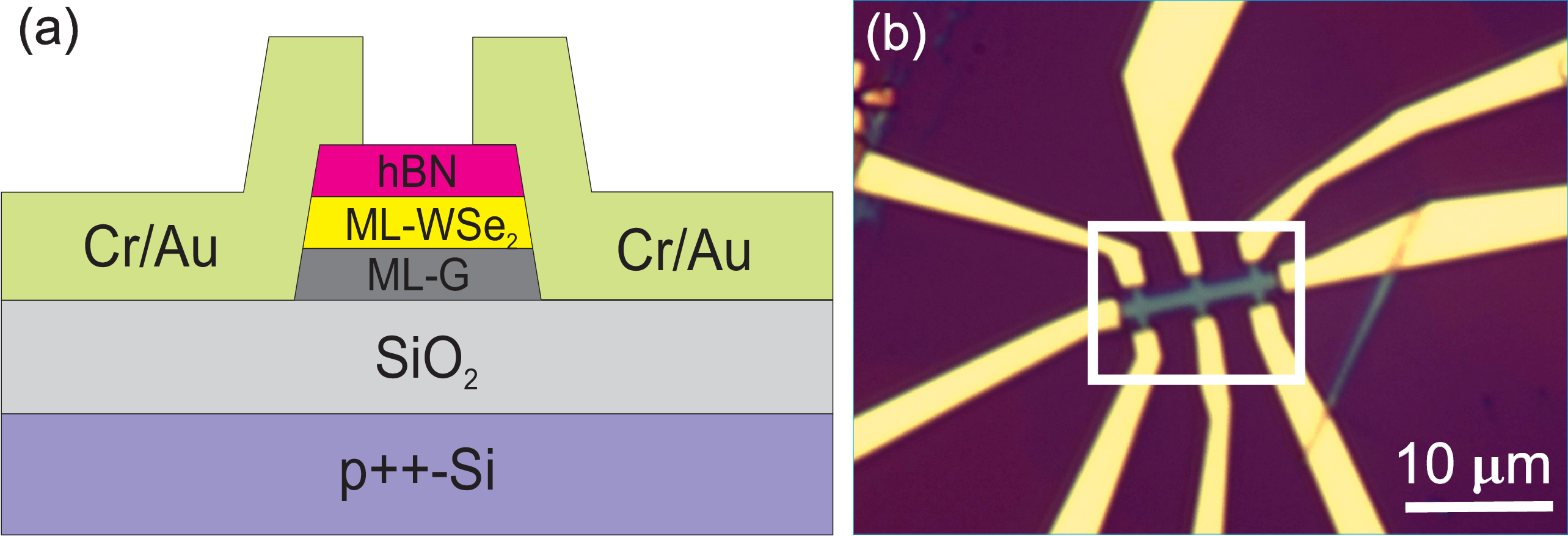}
    \caption{(a) ML-G/ML-WSe$_{2}$/hBN heterostructure with Cr/Au edge contacts on a Si/SiO$_{2}$ substrate. (b) Optical microscope image of sample 6. The heterostructure is etched in a Hall-bar geometry and contacted with Cr/Au edge contacts. The white box highlights the Hall bar.}\label{Fig. 1}
\end{figure}

We prepared two types of ML-G/ML-WSe$_{2}$/hBN heterostructures to explore the rotation angle dependence of SOC. For type (a) (samples 1-3), ML-WSe$_{2}$ was exfoliated from commercially available bulk crystals, while in type (b) (samples 4-6), ML-WSe$_{2}$ grown by chemical vapor deposition (CVD) at Jena University \cite{George_2019,Turchanin2025} was used instead. 
Furthermore, for type (a) samples, only the fractured crystal edges were used for angular alignment, whereas type (b) samples also involved crystallographic etching to identify the zigzag direction. ML-G and hBN were exfoliated from bulk crystals of natural graphite, and hBN, which is provided by NIMS Japan, respectively.  The hBN flake was mainly used to protect the heterostructure. 
A schematic of the heterostucture is shown in Fig. \ref{Fig. 1} (a). The devices were fabricated with the hot pickup method using a polydimethylsiloxane stamp and a polycarbonate film \cite{Pizzocchero}. The heterostructure was deposited on a $p^{++}$-doped Si/SiO$_{2}$ substrate. The $p^{++}$-doped Si layer later served as a gate electrode. To obtain clean interfaces between the flakes of the heterostructure, a cleaning procedure using contact mode atomic force microscope (AFM) similar to the method in \cite{KimY., Rosenberger} was applied. Afterwards, electron beam lithography (EBL) and reactive ion etching with CHF$_{3}$/O$_{2}$ gas were carried out to shape the heterostructure into a Hall-bar, as shown in Fig.~\ref{Fig. 1} (b) \cite{LWang}. Finally, edge contacts \cite{LWang} made of $0.5\,$nm chromium (Cr) and $100\,$nm gold (Au) were evaporated using physical vapor deposition (see Fig. \ref{Fig. 1}). The angular orientation between the respective crystal lattices was determined by two methods, which we describe below.

\subsubsection{\label{subsubsec:level2}Twist angle determination for sample type (a)\protect\\}

\begin{figure}[!h] 
		\centering
		\includegraphics[width=\linewidth]{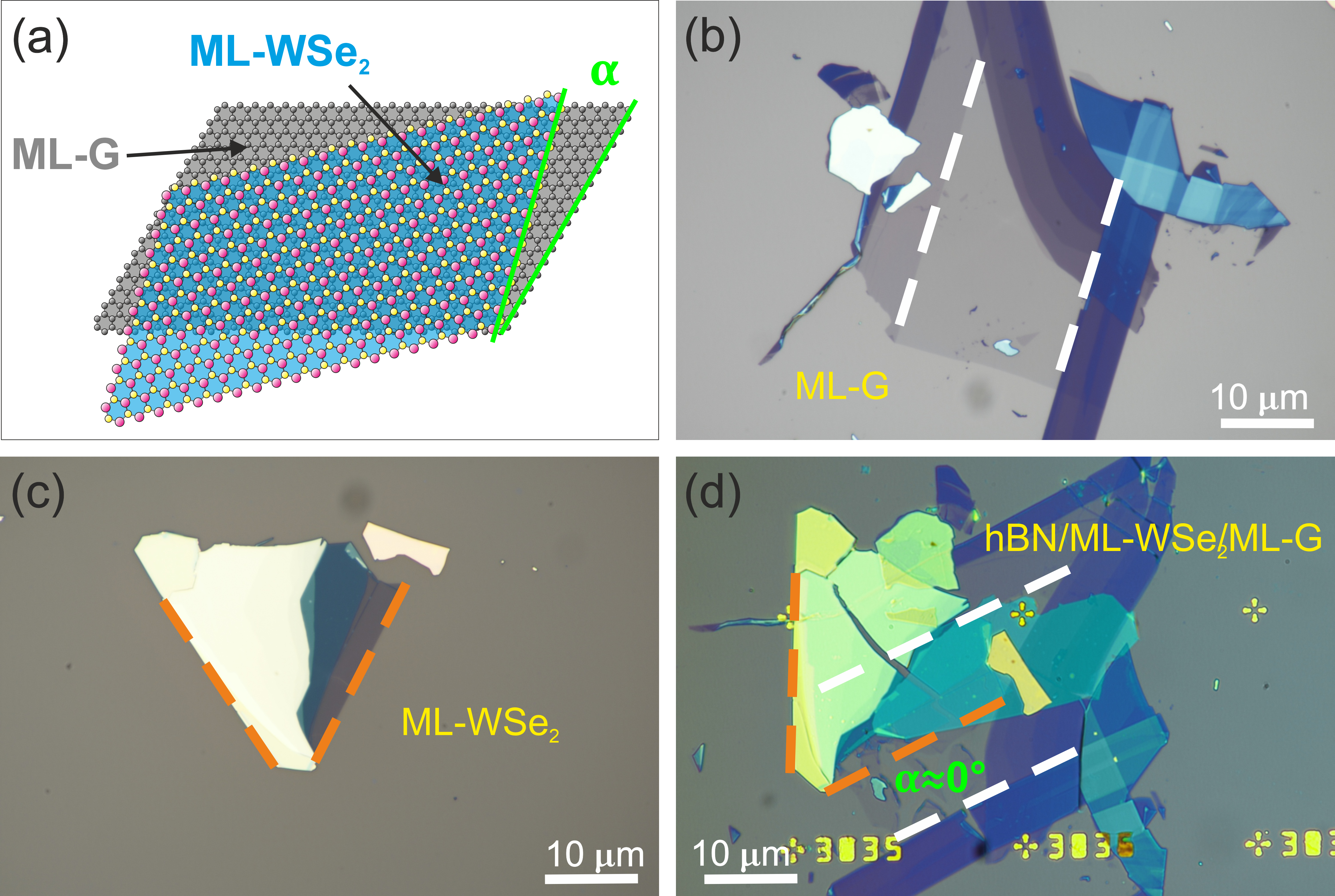}
		\caption{(a) Schematic representation of a ML-WSe$_{2}$ flake (blue) placed on a ML-G flake (gray). The lattices of the two flakes are rotated by the angle $\alpha$. (b) Graphite flake with a part of ML-G. The two white parallel lines mark specific edges, which are most likely of {zigzag} or {armchair} type. (c) WSe$_{2}$ flake is shown, part of which consists of ML-WSe$_{2}$. The orange lines here also mark specific edges with an angle of $60^\circ$ to each other. (d) Finished heterostructure. The ML-G and ML-WSe$_{2}$ flakes were superimposed in such a way that the specific edges enclose an angle of $0^\circ$. However, since the structure of the edges ({zigzag} or {armchair}) is not known, two possible rotation angles between the two lattices, $\alpha=0^\circ$ and $\alpha^\prime=30^\circ$, must be assumed here.}\label{Fig. 2}
\end{figure}

If two superimposed hexagonal crystal lattices are rotated with respect to each other (see Fig.~\ref{Fig. 2} (a)), a moiré pattern appears, and the resulting heterostructure has a $C_3$ ($120^\circ$) symmetry \cite{YangLi}. The range of twist angles yielding unique values for proximity-induced SOC, apart from a possible sign change of the SOC parameters, is limited to $\alpha=0^\circ...\,30^\circ$ \cite{YangLi}.
For the purposes of this study, it is therefore sufficient to focus on an angular range between $0^\circ$ and $30^\circ$ to determine the angular orientation of hexagonal 2D materials \cite{NaimerT,ZollnerK,YangLi}.\\
According to Ref.~\cite{Guo}, hexagonal 2D materials such as graphene, TMDCs and hBN tend to fracture during exfoliation with their edges aligned along the zigzag or armchair direction, where TMDCs in particular preferentially expose zigzag edges. Since the edge type of graphene cannot be distinguished from an optical micrograph, aligning fractured edges of graphene and TMDCs during stacking does not uniquely determine the twist angle.

 Therefore, for a measured angle of $\alpha$, extracted from the optical micrographs, the true twist angle could correspond to either $\alpha$ or $\alpha^\prime=30^\circ-\alpha$, unless $\alpha=15^\circ$, which is unambiguous.
To investigate the twist angle dependence of the SOC, the angle between the ML-G and the ML-WSe$_{2}$ was therefore set to approximately $15^\circ$ for sample 2 to ensure uniqueness. 
As for samples 1 and 3, angles of approximately $0^\circ$ or $30^\circ$ were selected, as very different and therefore very characteristic values of the SOC parameters were expected for the respective angles \cite{NaimerT,ZollnerK,YangLi}. Despite its relative simplicity, this method allows us to demonstrate both the angle dependence of the SOC and the reproducibility of the samples.
Figs. \ref{Fig. 2} (b) and (c) show ML-G and ML-WSe$_2$ flakes which were used for fabricating the heterostructure in Fig.~\ref{Fig. 2} (d).
An angle of $\alpha=0^\circ$ was set between ML-G and ML-WSe$_{2}$. Since {zigzag} and {armchair} cannot be distinguished, the angle can also be $\alpha^\prime=30^\circ$.
\\
\subsubsection{\label{subsubsec:level3}Twist angle determination for sample type (b)\protect\\}

\begin{figure}[!h]
		\centering
		\includegraphics[width=\linewidth]{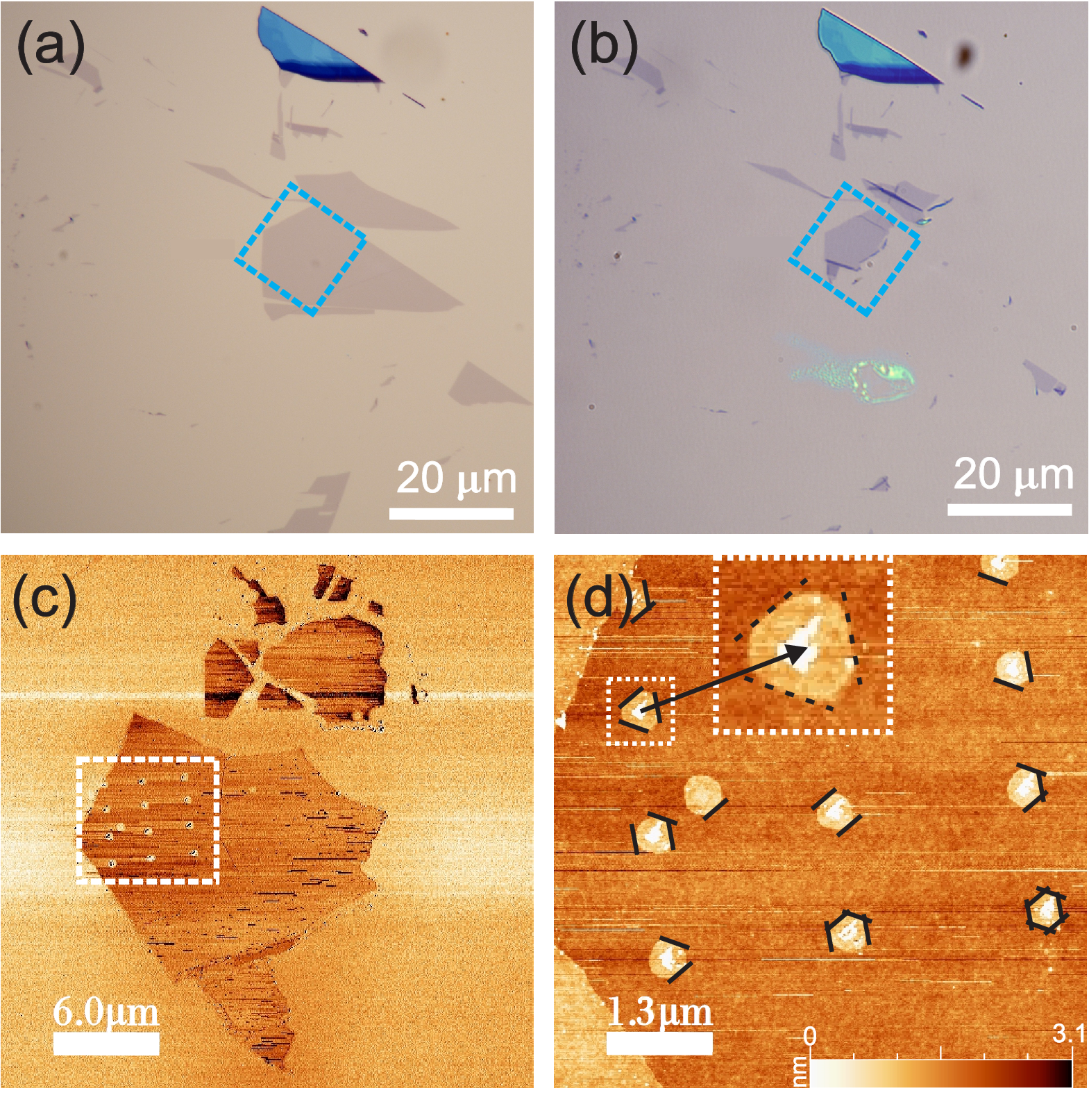}
		\caption{Different stages of a ML-G flake  before and after anisotropic etching. (a) ML-G/graphite flakes in their original shape after exfoliation. (b) remaining parts of the flakes after the other part has been removed to form the heterostructure.  (c) AFM z-topography of the anisotropically etched flakes (approximate position: blue frame in panels (a) and (b)). (d) Enlarged view of the area outlined in white showing anisotropically etched holes. The crystal orientation at sharp edges of the holes (black lines) becomes apparent.} \label{Fig. 3}
\end{figure}

For the type (b) samples, we developed a new method to determine the twist angle between ML-G and ML-WSe$_{2}$ unambiguously.
A CVD grown ML-WSe$_{2}$ was used, which forms zigzag edges during growth \cite{WeiFu,WangShanshan}, and we
identified the zigzag orientation of graphene by crystallographic etching \cite{Nemes-Incze,Oberhuber1}. 
For this purpose, several holes were first poked into the graphene flake using an AFM 
with a diamond-like carbon coated tip. The flake was then anisotropically etched in a CVD furnace with Ar/O$_{2}$ gas \cite{Nemes-Incze,Oberhuber1}, producing hexagonal holes bounded by zigzag edges, which allowed us to uniquely identify the crystallographic axes. For practical reasons, only part of the graphene flake (blue frame in Fig.~\ref{Fig. 3}) was subjected to high-temperature etching, while the remaining portion was mechanically separated during the tear-and-stack procedure \cite{YuanCao} and then used to assemble 
the heterostructure.
By comparing the AFM images (shown in Fig.~\ref{Fig. 3} (d)) with the optical micrographs, 
we can retrospectively determine
the crystal orientation of the ML-G flake, and consequently the twist angle in the heterostructure. For this study, we prepared three samples with twist angles  $\alpha=11^\circ$ (sample 4), $\alpha=22^\circ$ (sample 5) and $\alpha=15^\circ$ (sample 6).
\\
\subsection{\label{subsec:level2}Characterization\protect\\}
\begin{figure}[h]
	\centering
	\includegraphics[width=8.5cm]{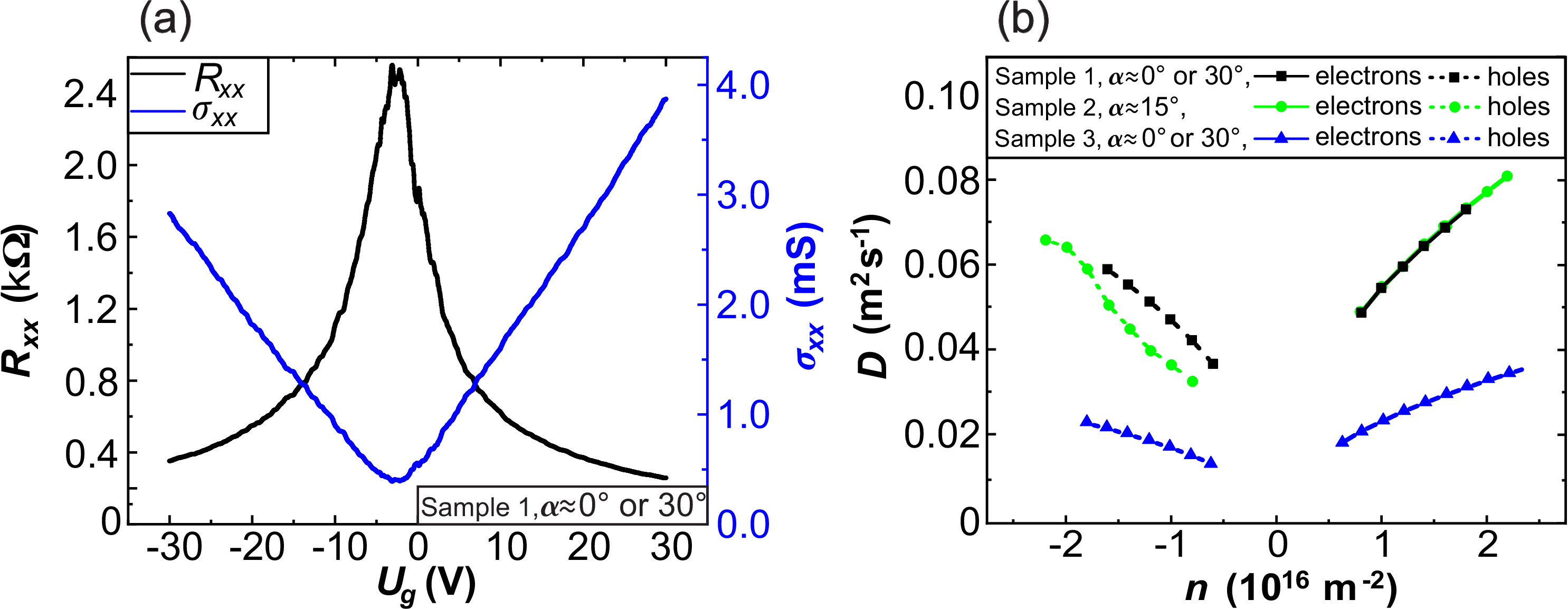}
    \caption{(a) Four-point resistance $R_{xx}$ and conductivity $\sigma_{xx}$ near the Dirac point for sample 1 as a function of gate voltage $U_{g}$. (b) Diffusion coefficient $D$ for samples 1-3 as a function of the charge carrier density $n$. 
    }\label{Fig. 4.}
\end{figure}

All measurements were carried out under cryogenic conditions at temperatures between $1.5$ and $1.7\,$K. An AC voltage $U_{bias}=0.3\,$V at a frequency $f=13~$Hz was applied to the Hall-bar in series with a bias resistor $R_{v}=10\,$M$\Omega$, delivering an AC current $I\approx 30\,$nA. Four-point lock-in measurements were carried out to obtain 
the four-point resistance $R_{xx}$ as a function of the gate voltage $U_{g}$ applied to the Si gate electrode of the Si/SiO$_{2}$ chip (see Figs. \ref{Fig. 4.} (a) and S1  (Supplementary Material) \cite{SM} for both $R_{xx}$ and $\sigma_{xx}$). Taking the sample dimensions into account, 
we determined the conductivity $\sigma_{xx}$ as a function of the gate voltage. From the slope of the $\sigma_{xx}$ vs. $U_{g}$ curve we determined the charge carrier mobility $\mu$ (see table \ref{Tabelle1}).
\begin{table}
    \centering
    \begin{tabular}{c|c|c|c|c|c}
       Sample & $\alpha$ &$L$ ($\mu$m) &$W$ ($\mu$m) & $\mu_{h}$ & $\mu_{e}$\\
       \hline
       1  & $0^\circ$ or $30^\circ$ & 4& 4& $\approx7800$& $\approx9300$\\
       2  & $15^\circ$ & 4& 4& $\approx6500$& $\approx9300$\\
       3  & $0^\circ$ or $30^\circ$ & 4& 4& $\approx2800$& $\approx4000$\\
       4  & $11^\circ$& 4& 1.5& $\approx6600$& $\approx5600$\\
       5  & $22^\circ$& 4& 1.5& $\approx3600$& $\approx2800$\\
       6  & $15^\circ$& 4& 1.5& $\approx6300$& $\approx7100$\\
    \end{tabular}
    \caption{Twist angle $\alpha$, Hall bar length ($L$) and width ($W$), and charge carrier mobilities $\mu_{h}$ (hole regime) and $\mu_{e}$ (electron regime) in cm$^{2}$/(Vs) for each sample.}
    \label{Tabelle1}
\end{table}
Using the mean free path  $l_{mfp} = \hbar\sqrt{\pi n}\mu/e$ we subsequently calculated the diffusion coefficient $D= v_F l_{mfp}/2$ and the momentum relaxation time $\tau_{p}=l_{mfp}/v_F$ as a function of the charge carrier density $n$ (see Figs. \ref{Fig. 4.} (b) and S2 (Supplementary Material) \cite{SM}).
\\
\subsection{\label{subsec:level2}Weak Anti-Localization and Fit\protect\\}

Localization effects arising from the interference of time-reversed paths are an important quantum mechanical correction to magnetotransport in the diffusive regime at low temperatures and low out-of-plane magnetic fields $B_{z}$. The phase coherence time $\tau_{\phi}$ of the system must be sufficiently long for the effect to be measurable. In particular, SOC leads to a weak antilocalization (SOC-WAL) signature around $B=0$. It provides an important means of determining the strength of the proximity-induced SOC \cite{ZheWang,Zihlmann,McCannFalko,Cummings}. 
In addition to SOC-WAL, localization effects can also manifest as weak localization (WL), or weak anti-localization arising from the Berry phase ($\pi$) in ML-G (Berry-WAL), depending on the strength of various scattering times \cite{McCannFalko,Zihlmann}. In addition, interference of electron paths leads to universal conductance fluctuations (UCF) \cite{Lundeberg}, obscuring the localization signals. UCFs can largely be removed from the data by gate averaging \cite{Gorbachev2007}, while the localization effects (SOC-WAL, WL and Berry-WAL) survive averaging and give rise to a quantum mechanical correction to the 
magnetoconductivity ($\Delta \sigma_{xx}(B)=\sigma_{xx}(B)-\sigma_{xx}(0)$)

A full expression including all relevant scattering times was given in Refs.~\cite{McCannFalko,Zihlmann}:
\begin{align}\label{WAL_groß}
	\begin{split}
		\Delta\sigma_{xx}(B_{z})=-\frac{e^{2}}{2\pi h}\left[ F\left(\frac{\tau_{B}^{-1}}{\tau_{\phi}^{-1}}\right) -F\left(\frac{\tau_{B}^{-1}}{\tau_{\phi}^{-1}+2\tau_{asy}^{-1}}\right)\right.\\
		\left. -2F\left(\frac{\tau_{B}^{-1}}{\tau_{\phi}^{-1}+ \tau_{asy}^{-1}+\tau_{sym}^{-1} }\right)
		-F\left(\frac{\tau_{B}^{-1}}{\tau_{\phi}^{-1}+2\tau_{iv}^{-1}}\right)\right.
		\\
		\left. -2F\left(\frac{\tau_{B}^{-1}}{\tau_{\phi}^{-1}+\tau_{*}^{-1}}\right)
		+F\left(\frac{\tau_{B}^{-1}}{\tau_{\phi}^{-1}+ 2\tau_{iv}^{-1}+2\tau_{asy}^{-1} }\right) \right.\\
		\left. +2F\left(\frac{\tau_{B}^{-1}}{\tau_{\phi}^{-1}+ \tau_{*}^{-1}+2\tau_{asy}^{-1} }\right)\right.\\
		\left.+2F\left(\frac{\tau_{B}^{-1}}{\tau_{\phi}^{-1}+ 2\tau_{iv}^{-1}+\tau_{asy}^{-1} +\tau_{sym}^{-1}}\right) \right.\\
		\left.+4F\left(\frac{\tau_{B}^{-1}}{\tau_{\phi}^{-1}+\tau_{*}^{-1}+\tau_{asy}^{-1}+\tau_{sym}^{-1}}\right)\right],
	\end{split}
\end{align}
with $\tau_{B}^{-1}=4eDB_{z}/\hbar$ and $F(z)=\ln(z) + \psi (0.5+z^{-1})$, where $\psi$ is the digamma function and $D$ is the diffusion coefficient. Important scattering times here are the phase coherence time $\tau_{\phi}$, the symmetric and asymmetric spin-orbit scattering times $\tau_{sym}$ and $\tau_{asy}$, which correspond to contributions to SOC that are symmetric or asymmetric under reversal of the $z$-coordinate \cite{McCannFalko}. 
Electron scattering within or between valleys is described by the intra- and intervalley scattering times $\tau_{intra}$ and $\tau_{iv}$, respectively. $\tau_{intra}$ is contained in $\tau_{*}^{-1}=\tau_{iv}^{-1}+\tau_{intra}^{-1}$ \cite{McCannFalko,Zihlmann}.\\

For short $\tau_{iv}$, the full expression in Eq.~\ref{WAL_groß} reduces to a simpler formula
given explicitly by McCann and Fal'ko \cite{McCannFalko}:
\begin{equation}\label{WAL_klein}
	\begin{split}
		\Delta\sigma_{xx}(B_{z})=-\frac{e^{2}}{2\pi h}\left[F\left(\frac{\tau_{B}^{-1}}{\tau_{\phi}^{-1}}\right)-F\left(\frac{\tau_{B}^{-1}}{\tau_{\phi}^{-1}+2\tau_{asy}^{-1}}\right)\right.\\
		\left.-2F\left(\frac{\tau_{B}^{-1}}{\tau_{\phi}^{-1}+\tau_{asy}^{-1}+\tau_{sym}^{-1}}\right)\right]
	\end{split}
\end{equation}
This expression was used by several experimental groups \cite{ZheWang,Voelkl,Wakamura} but it is not applicable here, as $\tau_{iv}$ is not vanishingly small in our experiments.

Fitting Eq.~\ref{WAL_groß} only results in a range of interdependent parameters. Therefore, in order to understand the significance and manifestation of each scattering time in the magnetotransport curves, we inspected Eq.~\ref{WAL_groß} numerically, starting from the results of a first fit and varying parameters systematically. We note the following observations: If the phase coherence time $\tau_{\phi}$ is reduced, e.g., if the temperature is increased, SOC-WAL disappears first, then WL and finally Berry-WAL, as expected according to \cite{ZheWang, Tikhonenko, Tikhonenko2}. The intervalley scattering time $\tau_{iv}$ and the intravalley scattering time $\tau_{intra}$ have a significant effect on WL and Berry-WAL \cite{Tikhonenko,Tikhonenko2}, but only limited influence on SOC-WAL. While Berry-WAL is mainly expected in very pure samples with weak elastic scattering ($\tau_{iv},\tau_{intra}  \gtrapprox  \tau_{\phi}$) \cite{Tikhonenko,Tikhonenko2},  WL dominates for relatively strong elastic scattering ($\tau_{iv},\tau_{intra}\ll\tau_{\phi}$) \cite{Tikhonenko,Tikhonenko2}. Finally, the scattering times $\tau_{asy}$ and $\tau_{sym}$ give rise to SOC-WAL \cite{McCannFalko}.  However, these two scattering parameters affect WL or Berry-WAL, prevalent at larger magnetic fields, less strongly.
These observations allow us to disentangle the contributions of those scattering times in our data.

In the samples of this work, Berry-WAL was generally dominant over WL in the individual measurements, presumably due to the relatively wide sample geometry. The resulting weak intervalley scattering precludes Eq.~\ref{WAL_klein} and necessitates Eq.~\ref{WAL_groß} for fitting. 
Fortunately, the problem of fitting a five parameter expression can be simplified considerably. In contrast to the other parameters, the phase coherence time $\tau_{\phi}$ mainly affects the height and width of the SOC-WAL peak near $B_{z}=0\,$T, so that $\tau_{\phi}$ can be approximated well in the fit. Alternatively, $\tau_{\phi}$ can still be determined from UCF using the autocorrelation function \cite{Lundeberg}, which generally allows $\tau_{\phi}$ to be determined quite accurately, leaving only four unknown fitting parameters. 
When examining the fit curve, we notice that  $\tau_{asy}$ and $\tau_{sym}$ mainly affect the low field behavior, while $\tau_{iv}$ and $\tau_{intra}$  predominantly act on the high field part, assisting us in selecting an acceptable parameter span. The possible fitting range can be narrowed down even further by assuming that the scattering rate $\tau_{SOC}^{-1}=\tau_{asy}^{-1}+\tau_{sym}^{-1}$  depends on the momentum relaxation time $\tau_{p}$ in a linear fashion \cite{Cummings}. Finally, 
the Rashba-SOC and valley-Zeeman-SOC calculated with \cite{Cummings}
\begin{equation}\label{eq. lambda_R}
\lambda_\mathrm{R}=\frac{\hbar}{\sqrt{4\cdot \tau_{asy}\cdot\tau_{p}}},
\end{equation}
\begin{equation}\label{eq. lambda_VZ}
\lambda_\mathrm{VZ}=\frac{\hbar}{\sqrt{4\cdot \tau_{sym}\cdot\tau_{iv}}}
\end{equation}
have to be independent on the charge carrier density $n$ \cite{McCannFalko},
which leads to further constraints on the fitting parameters and can be verified in the fitting results (cf. Figs.~\ref{Fig. 7.} and \ref{Figure_17})
\begin{figure}[!h]
		\centering
		\includegraphics[width=7cm]{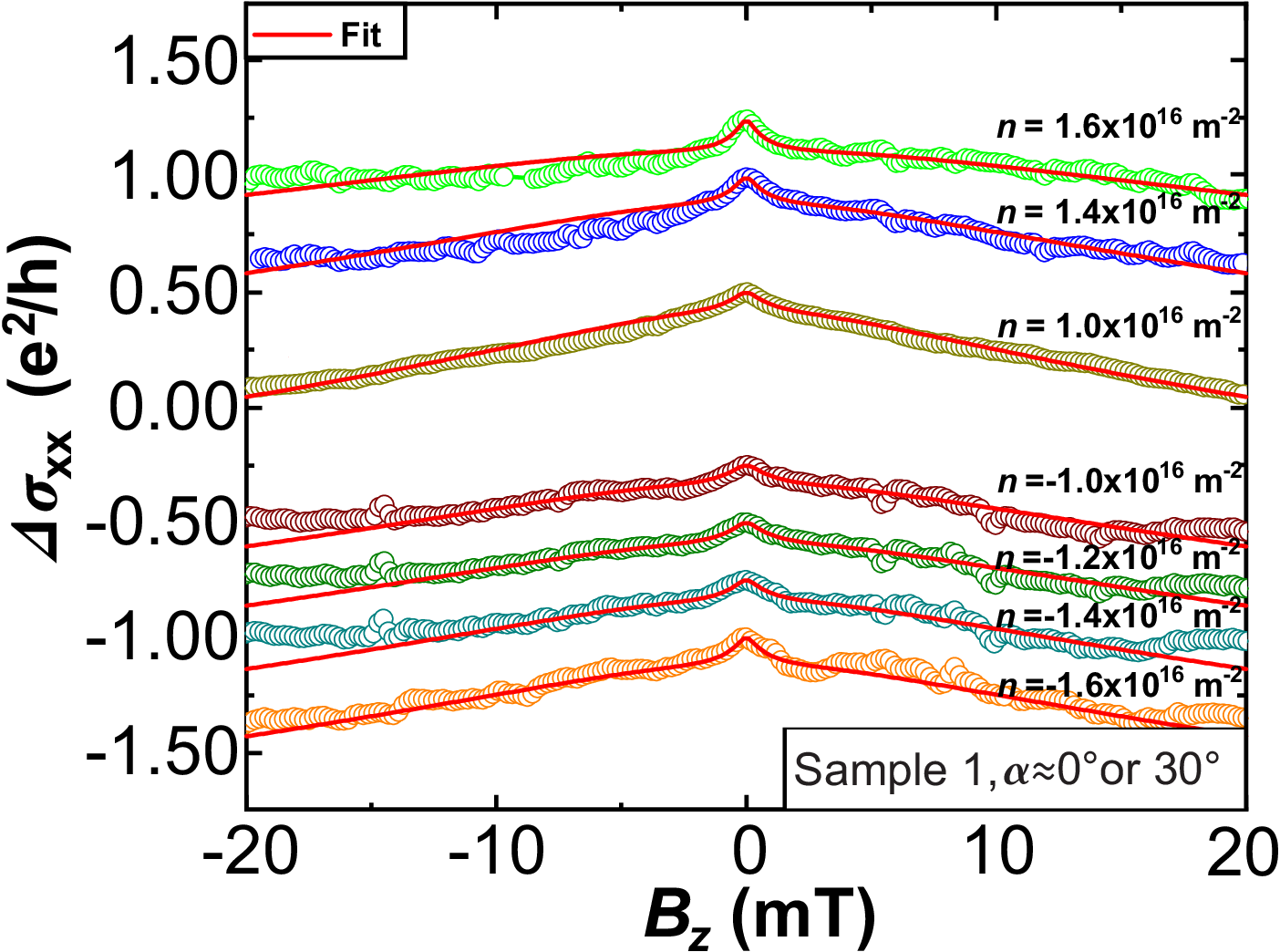}
		\caption{Measurements of the quantum mechanical correction of the conductivity $\Delta\sigma_{xx}$ for different charge carrier densities $n$ as a function of the out-of-plane magnetic field $B_{z}$. Curves are offset for clarity. 
		SOC-WAL is mainly responsible for the small significant peak near $B_{z}=0\,$T, while Berry-WAL, results in the decreasing conductivity towards larger magnetic fields. WL is not detected here due to the dominant Berry-WAL. Red lines: Fits to Eq.~\ref{WAL_groß}
		with fitting parameters given in Fig. \ref{Fig. 6.}.} \label{Fig. 5.}
\end{figure}\\
Curve fitting was now possible with the remaining permitted values of the fit parameters. In Fig. \ref{Fig. 5.} the measured quantum mechanical corrections after gate averaging \cite{Gorbachev2007} and corresponding fit curves are shown for sample 1 (for samples 2-6, see Supplementary Material \cite{SM} Fig. S3). The fit curves shown do not represent the only solutions. Instead, the parameters of the curves lie within a permitted value range of the respective fitting parameters, which are shown in Fig. \ref{Fig. 6.} for samples 1-3 and Fig. S4 (Supplementary Material \cite{SM}) for samples 4-6.
\begin{figure}[h]
		\centering
		\includegraphics[width=8.5cm]{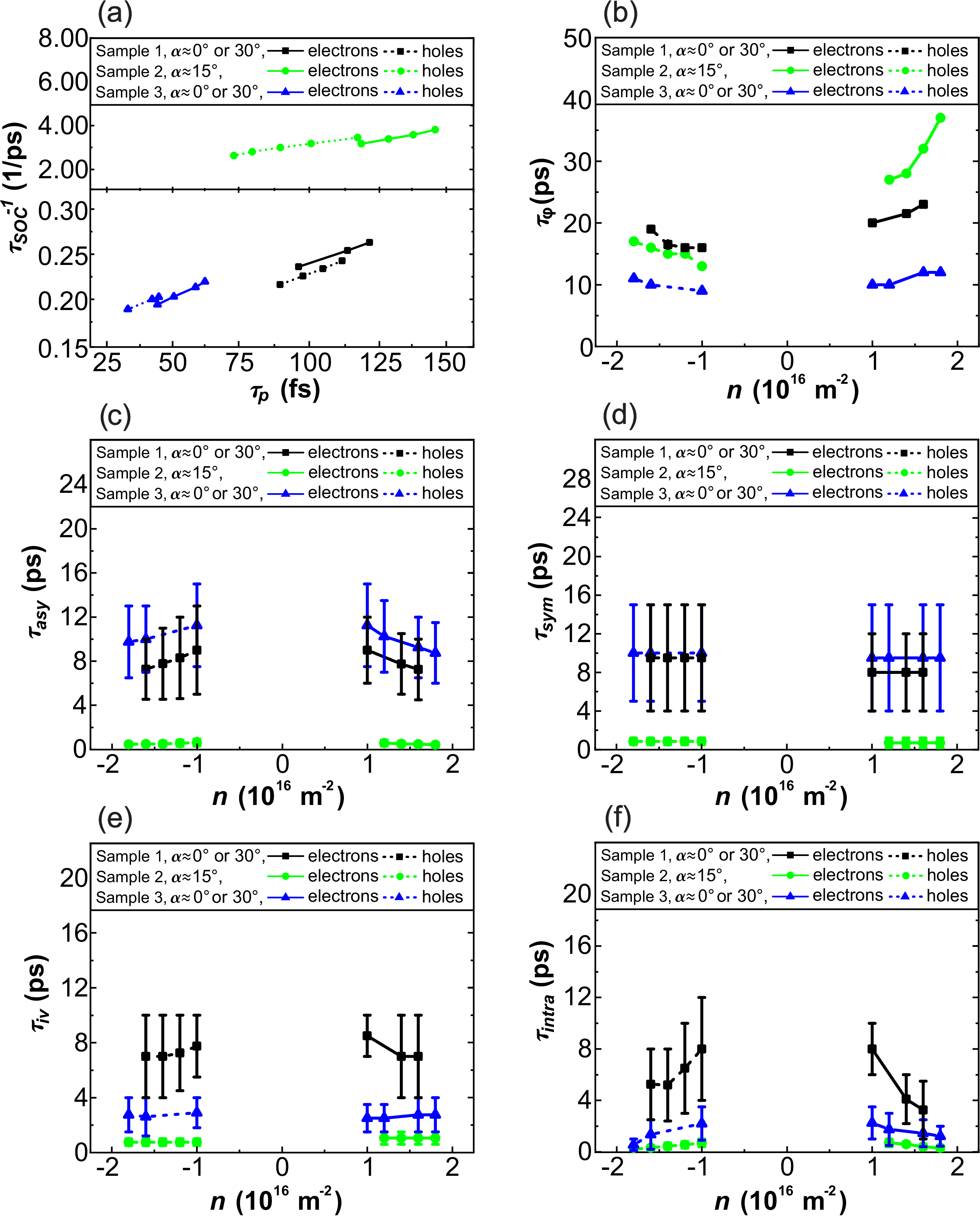}
    \caption{(a) $\tau_{SOC}^{-1}$ calculated from the average values of $\tau_{asy}$ and $\tau_{sym}$ as a function of the momentum relaxation time $\tau_{p}$ of samples 1-3 in the electron and hole regimes. The linear progression of the curves and the similar slope of the curves for both regimes are particularly striking. 
    (b) Phase coherence time $\tau_{\phi}$ as a function of the charge carrier density $n$. $\tau_{\phi}$ always increases with increasing $n$. (c)-(d) $\tau_{asy}$ and $\tau_{sym}$ of samples 1-3 for different $n$. $\tau_{asy}$ shows a dependence on $n$, whereas $\tau_{sym}$ does not. First signs of an angular dependence can also be seen for both parameters. (e)-(f) Intervalley scattering time $\tau_{iv}$ and the intravalley scattering time $\tau_{intra}$ of samples 1-3 are shown as a function of $n$. $\tau_{iv}$ shows no dependence on $n$. $\tau_{intra}$, on the other hand, decreases with increasing $n$. A systematic angle dependence is not apparent. 
    }\label{Fig. 6.}
\end{figure}\\
In Fig. \ref{Fig. 6.} (a), the linear dependence of $\tau_{SOC}^{-1}$ on the momentum relaxation time $\tau_{p}$  \cite{Cummings} is also shown. 
The slope of the straight line seems to remain constant even for electron and hole conduction. The phase coherence times $\tau_{\phi}$ as a function of the charge carrier density $n$ are depicted in Fig. \ref{Fig. 6.} (b). Similar to the findings in \cite{Tikhonenko,Tikhonenko2}, $\tau_{\phi}$ increases with increasing $n$.\\
In Fig. \ref{Fig. 6.} (c)-(d), we show the two spin-orbit scattering times $\tau_{asy}$ and $\tau_{sym}$ as a function of the charge carrier density $n$. It can clearly be seen here that $\tau_{asy}$ decreases with increasing density $n$, while $\tau_{sym}$ is independent on the density.  
Since $\tau_{sym}$ remains virtually constant here, the slope of $\tau_{SOC}^{-1}$ in Fig. \ref{Fig. 6.} (a) is strongly related to the slope of $\tau_{asy}$. In Fig. \ref{Fig. 6.} (e)-(f), the intervalley scattering time $\tau_{iv}$ and the intravalley scattering time $\tau_{intra}$ are shown as a function of the charge carrier density $n$. While $\tau_{iv}$ shows no significant dependence on the density $n$, $\tau_{intra}$ drops sharply with increasing density $n$, similar to the observations in \cite{Tikhonenko}.
The error bars represent the permitted range to yield acceptable fits in the procedure outlined above.
\subsection{\label{sec:level2}Twist angle dependence of SOC\protect\\}
By applying equations (\ref{eq. lambda_R}) and (\ref{eq. lambda_VZ}) and inserting the corresponding scattering times, we determined the two SOC parameters $\lambda_\mathrm{R}$ and $\lambda_\mathrm{VZ}$.
\begin{figure}[h]
	\centering
	\includegraphics[width=8.5cm]{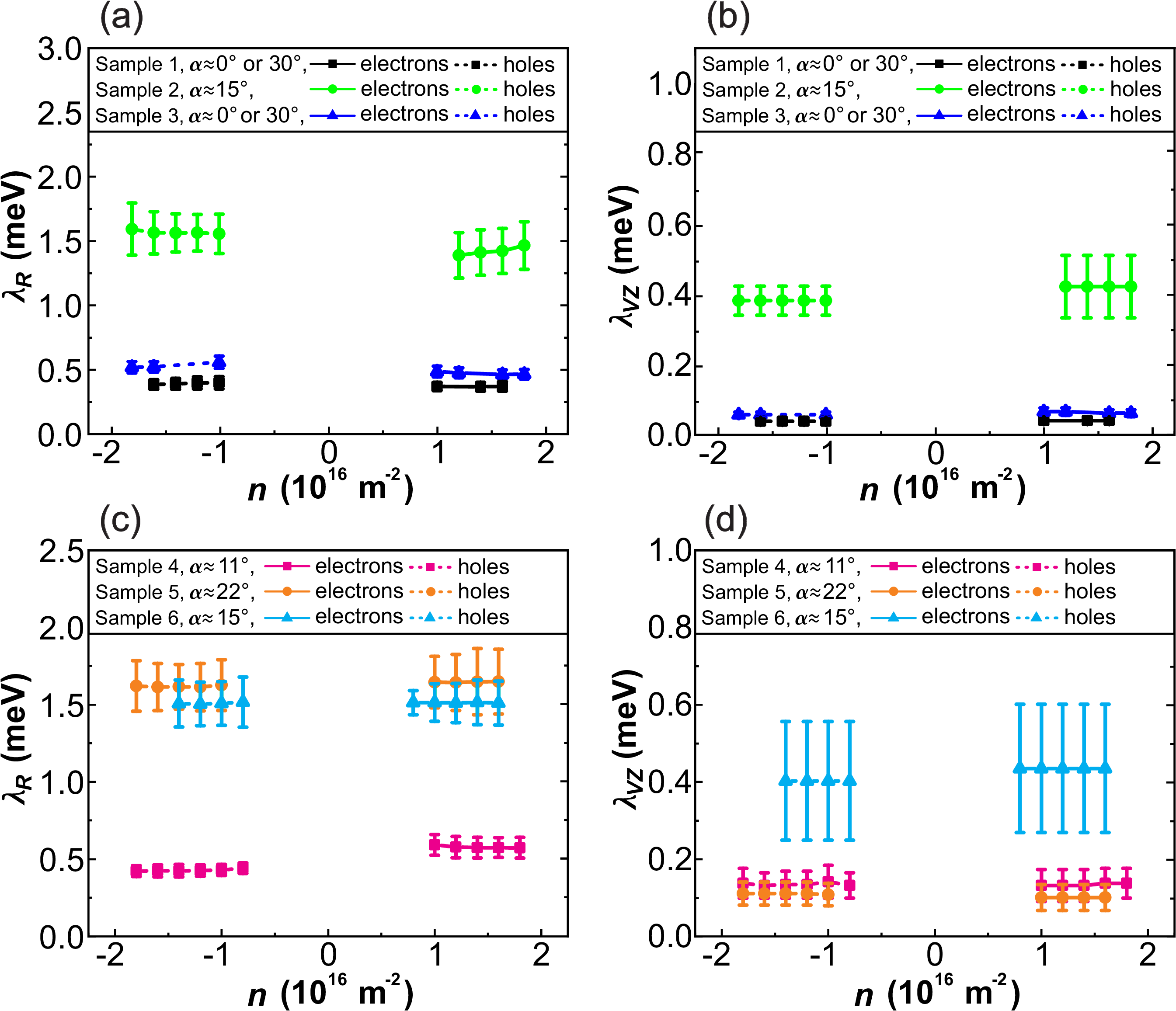}
    \caption{(a)-(b) $\lambda_\mathrm{R}$ and $\lambda_\mathrm{VZ}$ for samples 1-3 (type (a)) as a function of the charge carrier density $n$. (c)-(d) $\lambda_\mathrm{R}$ and $\lambda_\mathrm{VZ}$ for samples 4-6 (type (b)) as a function of the charge carrier density $n$.}\label{Fig. 7.}
\end{figure}
The results for type (a) samples 1-3 are shown in Fig. \ref{Fig. 7.} (a) and (b). Here it can be seen that the two SOC parameters for sample 2 with a defined angle of $\alpha=15^\circ$ are much larger than the SOC parameters for the other two samples, which have angles of $\alpha=0^\circ$ or $\alpha=30^\circ$. This already indicates an angle dependence for both the Rashba-SOC and the valley-Zeeman-SOC. Comparing samples 1 and 3, we also notice that they have nearly identical values for both $\lambda_\mathrm{R}$ and $\lambda_\mathrm{VZ}$. This suggests that both samples may have nearly the same twist angle and that the results are reproducible. 
Considering the results for $\lambda_\mathrm{R}$ and $\lambda_\mathrm{VZ}$ of type (b) samples 4-6, shown in Fig. \ref{Fig. 7.} (c)-(d), we note that different values of the SOC parameters are associated with different angles. Such a variation would also be expected if the sample preparation and evaluation were not reproducible. However, if we compare $\lambda_\mathrm{R}$ and $\lambda_\mathrm{VZ}$ for samples 2 and 6, both with $\alpha=15^\circ$ (which is unambiguously defined in both fabrication methods), we note that the parameter values are well reproduced. 
We thus conclude that identically defined twist angles also lead to similar SOC parameters, but different twist angles lead to a significant change in the SOC parameters. 
In this context, we not only observe a clear dependence of the proximity-induced SOC on the rotation angle but 
can also explain the poor reproducibility of SOC parameters when the twist angle is not controlled.
\subsection{\label{sec:level1}Discussion\protect\\}

\begin{figure}[h]
    \centering
    \includegraphics[width=8.6cm]{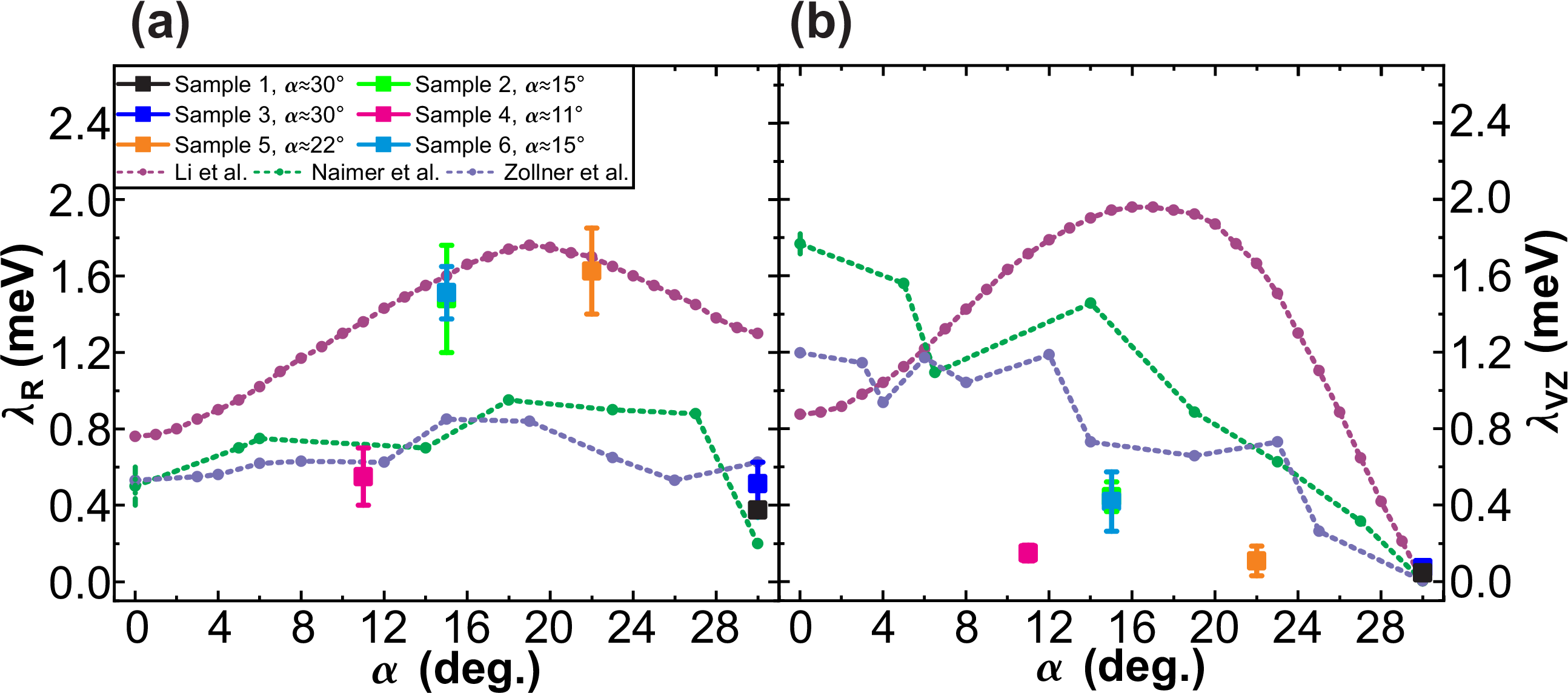}
    \caption{(a) Values of $\lambda_\mathrm{R}$ of samples 1-6 are shown graphically with corresponding rotation angles. In addition, the predicted values for $\lambda_\mathrm{R}$ from Li et al. \cite{YangLi}, Naimer et al. \cite{NaimerT} and Zollner et al. \cite{ZollnerK} are shown. (b) Similar representation for $\lambda_\mathrm{VZ}$. In (a) and (b) the green curve shows an error bar at $0^\circ$. Similarities and discrepancies between theory and experiment are discussed in the text.}\label{Fig. 8.}
\end{figure}

At this point, a comparison of the experimental findings with the predictions from theory \cite{YangLi,NaimerT,ZollnerK} can be carried out (see Fig. \ref{Fig. 8.}). Ref.~\cite{DavidA} was not taken into account, as no specific calculations on ML-WSe$_{2}$ were provided there. First of all, a comparison of Refs.~\cite{YangLi,NaimerT,ZollnerK} shows that, while the qualitative trend is similar across different theoretical methods, there are some quantitative discrepancies.
Now we turn to the experimental data. First, to resolve the possible ambiguity of the twist angle in samples 1 and 3, we note that $\lambda_\mathrm{VZ}\approx 0.05-0.06\,$meV  would clearly not match any theoretical prediction for $\alpha=0^\circ$. We therefore conclude that the true twist angle for both samples is $\alpha=30^\circ$, where $\lambda_\mathrm{VZ}=0$ was predicted by all theories and experiment confirms that valley-Zeeman-SOC can actually be switched off.

For the Rashba-SOC parameter $\lambda_\mathrm{R}$, we find good agreement in both trend and quantitative results with theoretical predictions, where experimental results better match the predictions of Ref.~\cite{YangLi} for samples 2, 5, and 6 and the other samples fall in line with the calculations in Refs.~\cite{NaimerT,ZollnerK}. The valley-Zeeman parameter $\lambda_\mathrm{VZ}$ does not quantitatively match theoretical predictions, with the possible exception of the $\alpha=15^\circ$ samples, which come close to the calculations in Ref.~\cite{ZollnerK}. We note that in the Supplemental Material to Ref.~\cite{Peterfalvi2022}, Péterfalvi {\em et al.} provide calculations for the MLG/WSe$_2$ case, where $\lambda_\mathrm{VZ}$ remains close to zero over a range of twist angles.
In summary, the experimental results show a clear dependence of the proximity-induced SOC on the twist angle. It has also been demonstrated that the samples exhibit a reproducible SOC when the rotation angle is set to the same value. In addition, it was confirmed that the valley-Zeeman-SOC can be eliminated by selecting an angle of $\alpha=30^\circ$. However, there are still discrepancies between theory and experiment.\\
The reason for this may possibly lie in the existing lattice strain. Each rotation angle is accompanied by a certain lattice strain, which results in a corresponding strength of the proximity-induced SOC \cite{YangLi, NaimerT, ZollnerK}. In practice, however, the lattice strain can also arise from other sources (e.g. defects, impurities...) unrelated to the rotation angle. This can result in deviations in the strength of the proximity-induced SOC when comparing theory and experiment.\\
Recently, a new theory of weak localization in graphene including SOC was presented \cite{Golub, Golub2026}, which predicts deviations from the theory by McCann and Fal'ko \cite{McCannFalko} whenever $\tau_{asy}$ is of the same order as $\tau_{\phi}$. 
While this theory can possibly account for the quantitative mismatch of some data points, establishing a detailed fitting procedure and extracting meaningful values from experimental data goes beyond the scope of this work.

\section{Tuning proximity-induced SOC using mechanical pressure\protect\\}\label{sec:3}

To further explore the tunability of the proximity-induced SOC, we also performed experiments on sample 1 under hydrostatic pressure, which affects 
the interlayer distance in heterostructures \cite{Gmitra1}, and hence results in an enhancement of the proximity induced SOC \cite{Balint,BalintPRB2025,Kedves2023}.
To apply hydrostatic pressure we used a pressure cell, described in detail in Ref.~\cite{BalintF}, which is filled with non-polar kerosene. Kerosene serves as a pressure transferring medium, and the cell is pressurized at room temperature, then cooled down to $T=1.5$ K. 
For a consistency check with the data described in section \ref{sec:2}, we first performed experiments without kerosene and at  $p=0\,$GPa, then, after warming up, filled the cell with kerosene, pressurized to $p=1.9\,$GPa and cooled down again for the second set of experiments.
The curve fitting and subsequent evaluation were carried out in the same way as the procedure described in the previous section.\\

\subsection{Characterization\protect\\}
\begin{figure}[!h]
 \begin{minipage}[h]{\linewidth} 
     \centering
     \includegraphics[width=8.5cm]{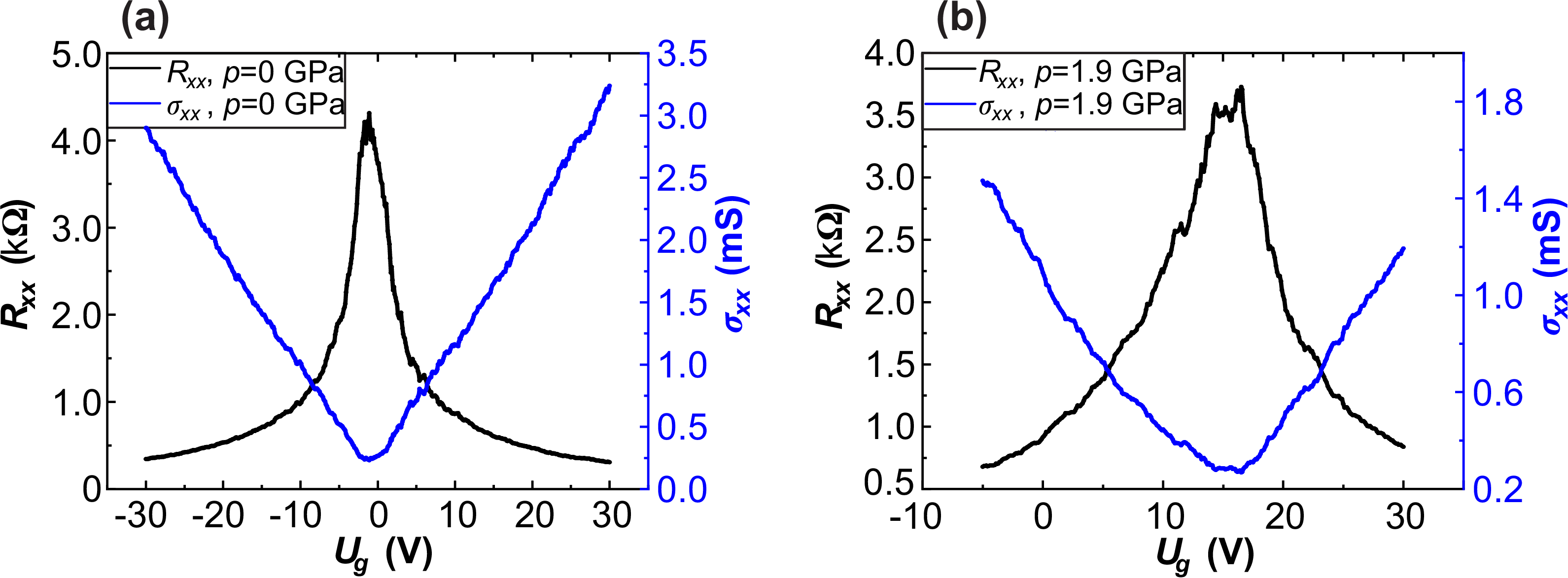}
 \end{minipage}
 \caption{Measurements of the gate electric field effect on resistance $R_{xx}$ and  conductivity $\sigma_{xx}$. (a) without pressure ($p=0\,$GPa) and without kerosene and (b) $p=1.9\,$GPa and with kerosene. The resistance peak for (b) is slightly wider than for (a) and also the Dirac point has shifted into the positive range by applying pressure.}\label{Figure_13}
\end{figure}

\begin{figure}[!h]
 \begin{minipage}[h]{\linewidth} 
     \centering
     \includegraphics[width=8.5cm]{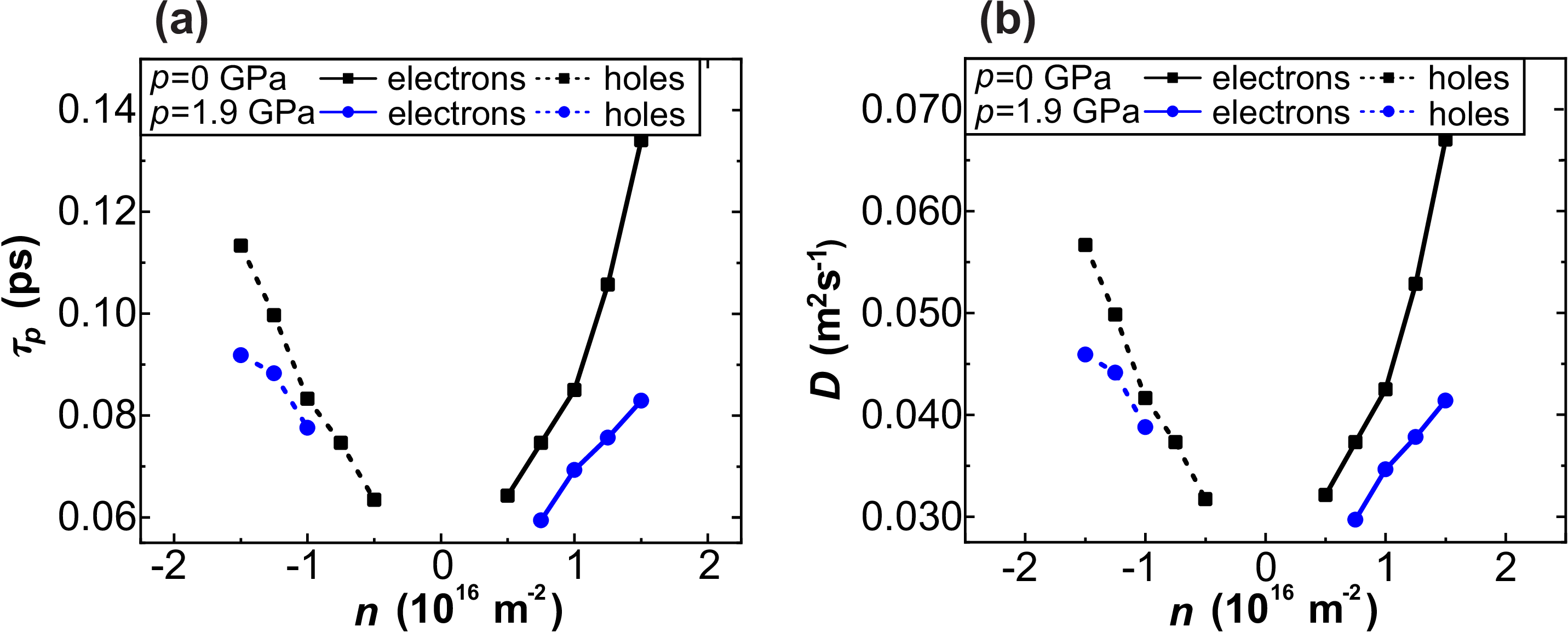}
 \end{minipage}
 \caption{(a) Momentum relaxation time $\tau_{p}$ and (b) diffusion coefficient $D$ for $p=0\,$GPa and $p=1.9\,$GPa as a function of the charge carrier density $n$.}\label{Figure_14}
\end{figure}

An AC voltage $U_{bias}=100\,\mu$V with a frequency $f=1137\,$Hz was applied to the sample, causing a current $I\approx10-50\,$nA to flow. Then the electric field effect measurements were conducted both with and without applied pressure (see Fig. \ref{Figure_13}).
\\
The curves of $R_{xx}$ and $\sigma_{xx}$ as a function of the gate voltage $U_{g}$ show the usual shape, allowing us to extract background doping and mobility. Both differ slightly between the sets of experiments, however, it is not clear if this is due to different pressure or the absence or presence of kerosene in the cell.
Using the charge carrier mobility $\mu$ ($p=0\,$GPa: $\mu_{e}\approx\mu_{h}\approx8000\,$cm$^{2}$V$^{-1}$s$^{-1}$, $p=1.9\,$GPa: $\mu_{e}\approx\mu_{h}\approx6000\,$cm$^{2}$V$^{-1}$s$^{-1}$), the momentum relaxation time $\tau_{p}$ and the diffusion coefficient $D$ were then determined (see Fig. \ref{Figure_14} (a)-(b)).
\\
There is a certain deviation in $\mu$ after kerosene and pressure were added, which is also accompanied by a corresponding change in $\tau_{p}$ and $D$.\\
\subsection{Weak Anti-Localization and Fit\protect\\}

\begin{figure}[!h]
 \begin{minipage}[h]{\linewidth} 
     \centering
     \includegraphics[width=8.5cm]{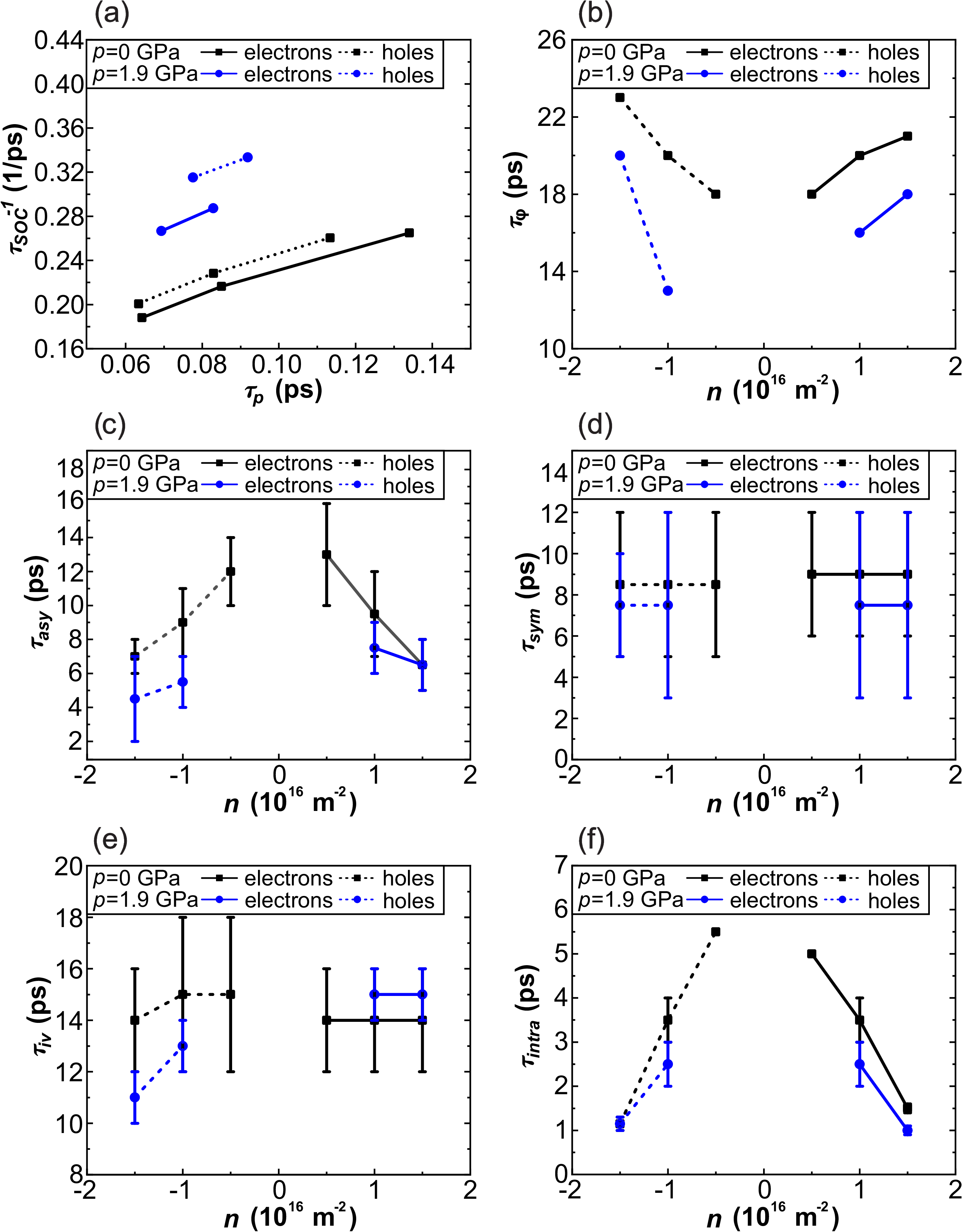}
 \end{minipage}
 \caption{(a) $\tau_{SOC}^{-1}$ as a function of $\tau_{p}$ for the two cases $p=0\,$GPa and $p=1.9\,$GPa. For the calculation of $\tau_{SOC}^{-1}$ the average values of $\tau_{asy}$ and $\tau_{sym}$ were used. (b)-(f) Progression of $\tau_{\phi}$, $\tau_{asy}$, $\tau_{sym}$, $\tau_{iv}$ and $\tau_{intra}$ as a function of the charge carrier density $n$ for the two cases $p=0\,$GPa and $p=1.9\,$GPa.}\label{Figure_16}
\end{figure}

We performed magnetoconductance measurements under hydrostatic pressure (data shown in the Supplementary Material \cite{SM}, Fig. S5), and the same analysis is made as described above. The results of the fitting procedure of Eq.~\ref{WAL_groß} on the SOC-WAL signal is shown in Fig.~\ref{Figure_16}.
\\
In Fig. \ref{Figure_16} (a) the scattering rate $\tau_{SOC}^{-1}$ as a function of $\tau_{p}$ again shows a linear dependence. Its slope is also almost identical for electrons and holes at both pressures. The phase coherence time  $\tau_{\phi}$ (see Fig. \ref{Figure_16} (b)) is found to be slightly reduced at high pressure, similar to earlier results \cite{Balint}, which suggest an increased number of scattering events when the heterostructure is compressed.
The asymmetric spin lifetime $\tau_{asy}$ (see Fig. \ref{Figure_16} (c)) becomes smaller with increasing pressure $p$ and increasing charge carrier density $n$, unlike $\tau_{sym}$ (see Fig. \ref{Figure_16} (d)), which, within the error of the extraction method, shows no dependence on the charge carrier density $n$ and no significant dependence on the pressure $p$. Both results are consistent with earlier observations  \cite{Balint}.
Finally, we observe no clear $p$ dependence of $\tau_{iv}$ and $\tau_{intra}$ (see Fig. \ref{Figure_16} (e,f)). A dependence on the charge carrier density $n$ can only be seen here for $\tau_{intra}$.\\
\subsection{Pressure dependence of SOC\protect\\}

\begin{figure}[!h]
    \begin{minipage}[h]{\linewidth} 
    \centering
   \includegraphics[width=8.5cm]{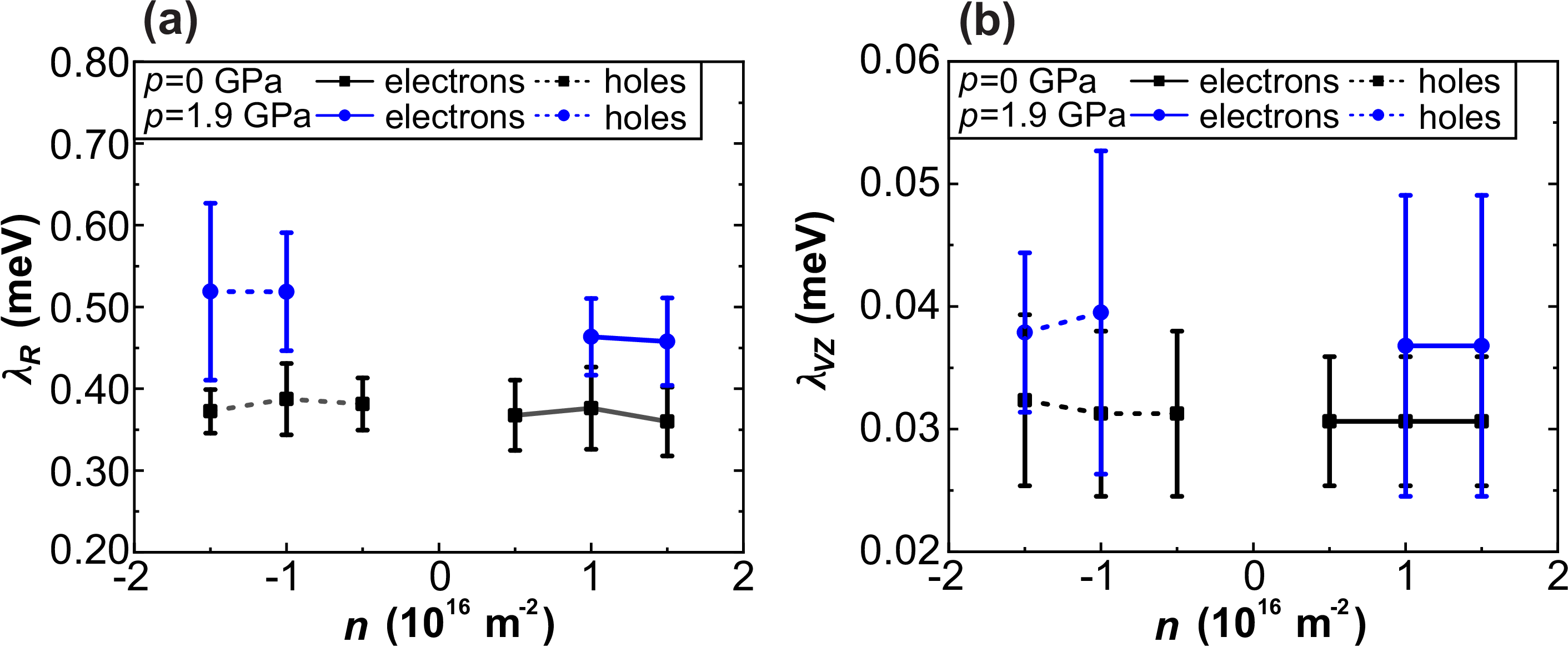}
     \end{minipage}
    \caption{SOC parameters $\lambda_\mathrm{R}$ (a) and $\lambda_\mathrm{VZ}$ (b) as a function of the charge carrier density $n$ at $p=0\,$GPa and $p=1.9\,$GPa.}\label{Figure_17}
\end{figure}

From the extracted scattering times, the SOC parameters $\lambda_\mathrm{R}$ and $\lambda_\mathrm{VZ}$ were determined as a function of $n$ using Eqs.~\ref{eq. lambda_R} and \ref{eq. lambda_VZ} (see Fig. \ref{Figure_17}). The Rashba-type SOC strength (panel a) is independent of $n$ as expected. However, a clear $p$ dependence is visible: at $p=1.9$ GPa, $\lambda_\mathrm{R}$ increased by ~40\%, which is consistent with previous findings \cite{Balint,BalintPRB2025}.
The valley-Zeeman-type SOC strength (panel b) is also independent of $n$, and a similar increase with $p$ can be observed. However, due to the larger error bars, this trend is less pronounced.

\section{Discussion and Conclusion\protect\\}
In the first part we showed that the twist angle between ML-G and ML-WSe$_2$ indeed determines the values for the SOC parameters $\lambda_\mathrm{R}$ and $\lambda_\mathrm{VZ}$, and obtained a good, albeit not perfect, match with numerical predictions. We also demonstrated that proximity-induced SOC becomes more reproducible when the twist angle is controlled during fabrication of van-der-Waals heterostructures. 

In the second part of this work, we performed magnetoconductance studies under hydrostatic pressure. By analyzing the WAL signals, we demonstrated a visible increase of the SOC parameters after applying pressure.  Our results confirm previous results \cite{Balint,Kedves2023,BalintPRB2025}.

To summarize, we have shown that SOC can be tuned both with the twist angle between crystal axes and the distance between the layers of graphene and WSe$_2$. Given that proximity-induced SOC in graphene is a necessary ingredient for advanced graphene-based spintronics \cite{Gmitra.2017}, spin quantum devices in bilayer graphene \cite{Dulisch2025,Gerber2025,Ge2025}, or the appearance of correlated states \cite{Zhang2023},
an in-depth understanding is essential for their controlled realization.

\begin{acknowledgments}
We thank L. E. Golub for helpful discussions on the theory of WAL. 
T.R., M.M., J.A., D.W., and J.E. acknowledge funding by the Deutsche Forschungsgemeinschaft (DFG, German Research Foundation) – Project-ID 314695032 – SFB 1277 (subproject A09) and 426094608 (ER 612/2).
B.S., S.C. and P.M. acknowledge support from the 2DSOTECH FlagERA network, by
the European Research Council ERC project Twistrain and funding from the European Innovation Council project 2DSPIN-TECH (No. 101135853),  from project  2024-1.2.10- TÉT-IPARI-IL-2024-00010 and also from the EKÖP-25-4-II-BME-46 University Research Scholarship Program of the Hungarian Ministry for Culture and Innovation from the source of the National Research, Development and Innovation Fund.
A.T. acknowledges funding by the Deutsche Forschungsgemeinschaft (DFG, German Research Foundation) – Project-ID 535253440 – SPP 2244 2DMP and Project-ID 398816777 – SFB 1375 NOA (subproject B2) as well as funding from the European Innovation Council project 2DSPIN-TECH (No. 101135853).
K.W. and T.T. acknowledge support from the JSPS KAKENHI (Grant Numbers 21H05233 and 23H02052) , the CREST (JPMJCR24A5), JST and World Premier International Research Center Initiative (WPI), MEXT, Japan.
\end{acknowledgments}

\section*{DATA AVAILABILITY}
The data that support the findings of this article are openly available~\cite{RockingerData}.


\providecommand{\noopsort}[1]{}\providecommand{\singleletter}[1]{#1}%

\foreach \x in {1,...,4}
{%
	\clearpage
	\includepdf[pages=\x]{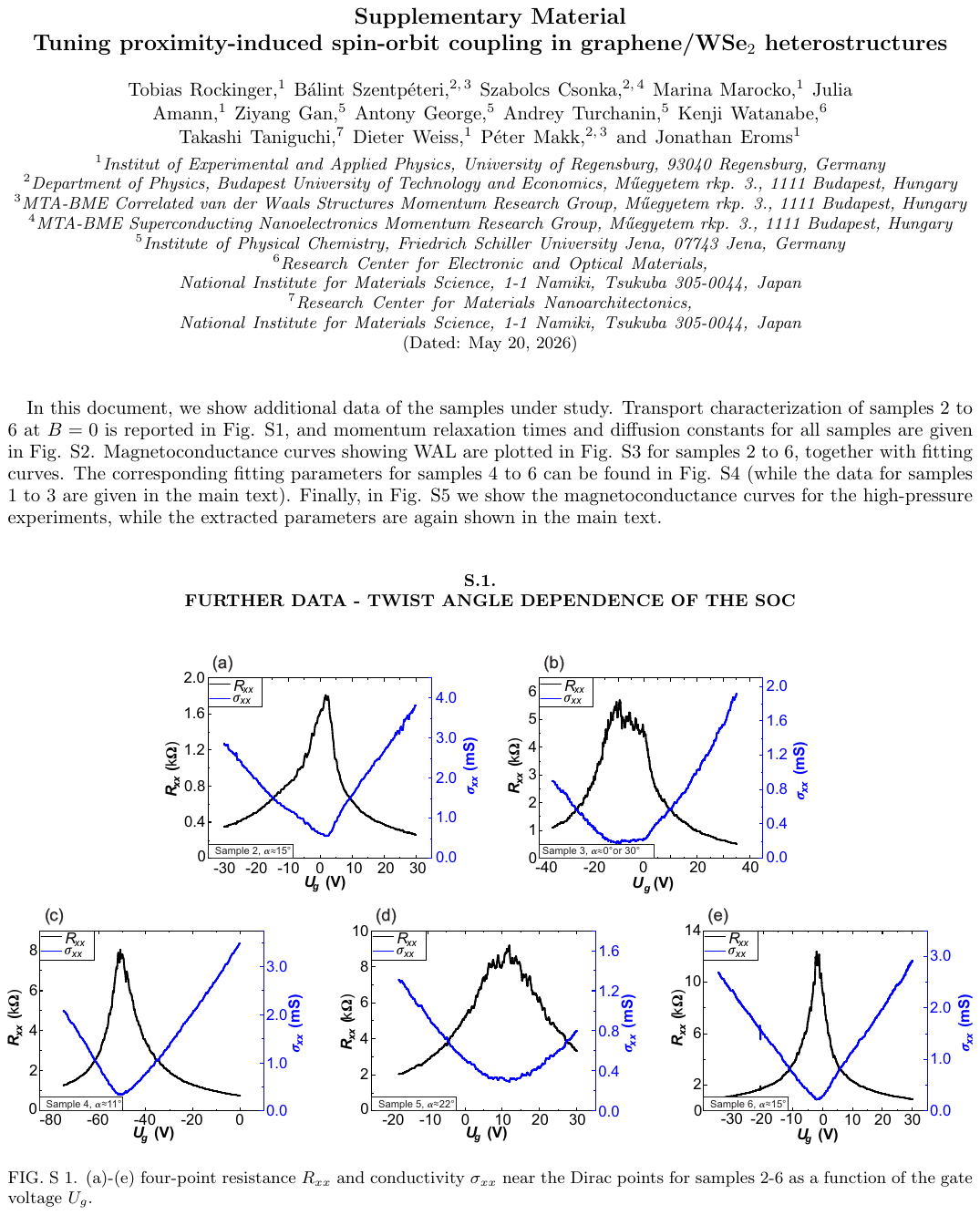}
}

\end{document}